\documentclass[10pt]{article}  

\pdfoutput=1

\usepackage{times}
\usepackage{graphicx}
\usepackage{standalone}
\usepackage[final]{pdfpages}
\usepackage[utf8]{inputenc}

\usepackage[a4paper,left=3cm,right=3cm,top=3.5cm,bottom=3.5cm]{geometry}

\begin{document}

\setcounter{page}{1}

\includepdf[pages=-]{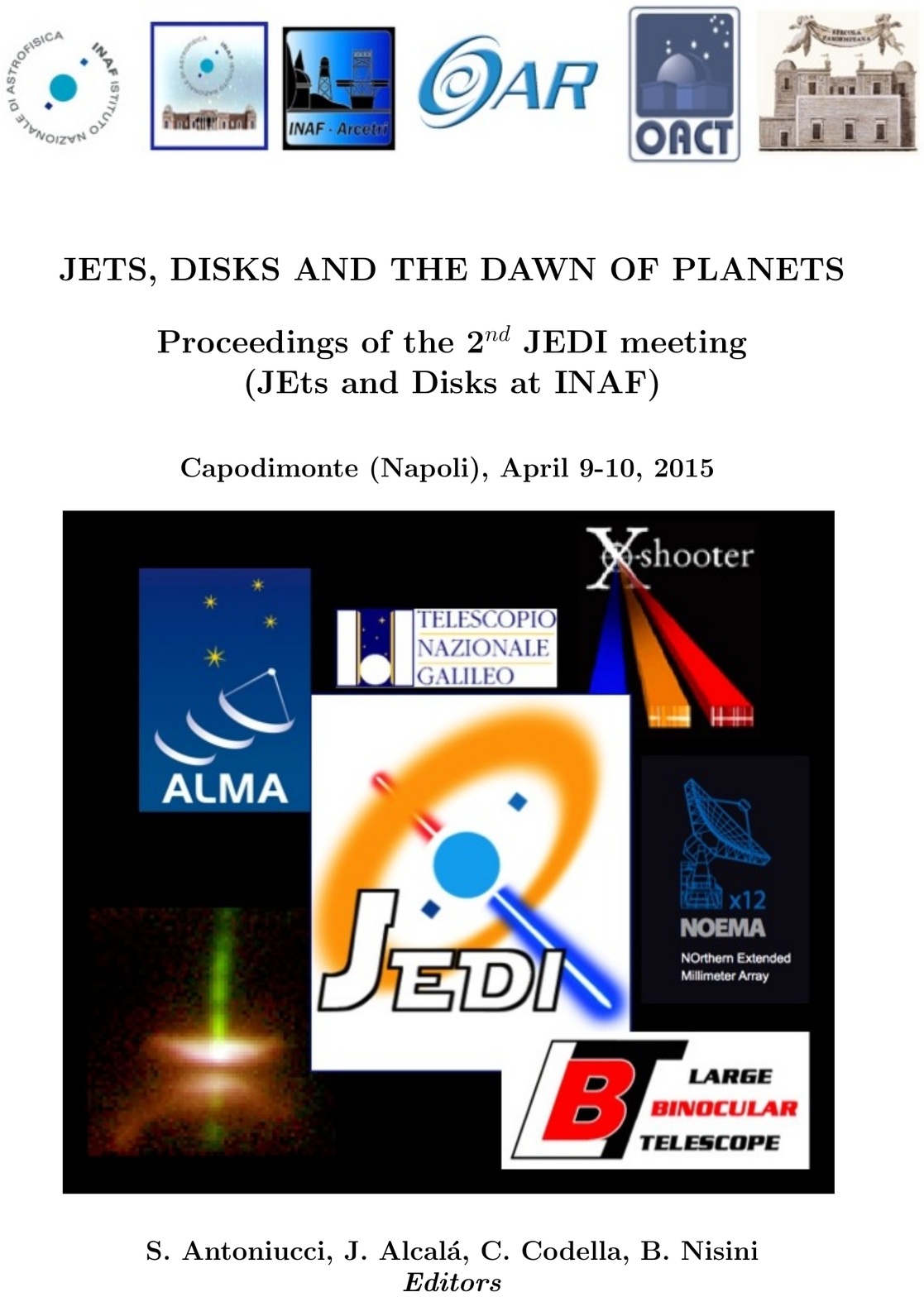}

\clearpage
\includepdf[pages=-,pagecommand={\thispagestyle{plain}}]{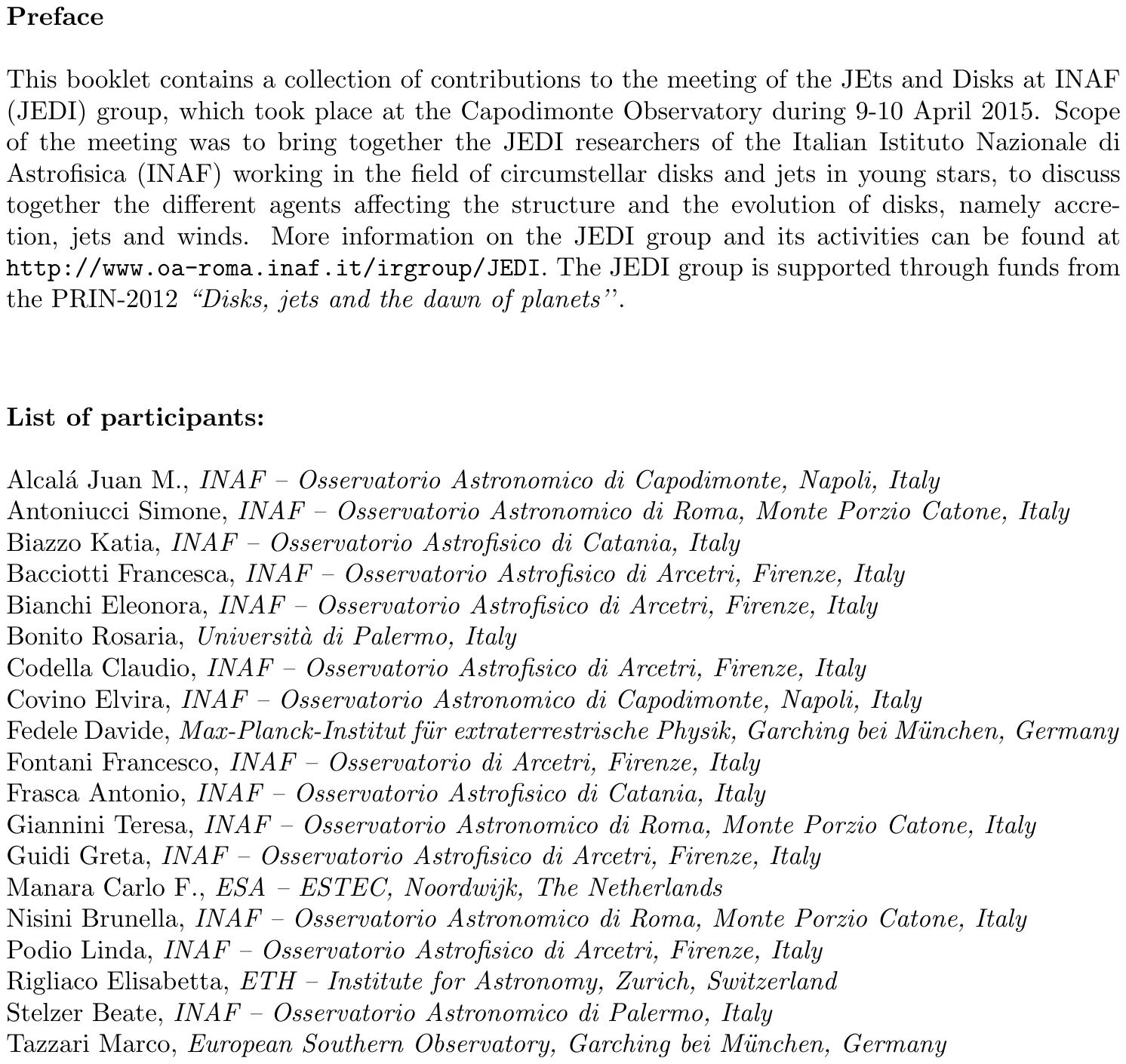}

\tableofcontents
\clearpage

\section*{\huge Part 1: \\~ \\ \textit{Early Protostellar Phases}}
\addcontentsline{toc}{section}{Part 1: Early Protostellar Phases}

\clearpage
\addcontentsline{toc}{subsection}{\textit{E. Bianchi} -- The CH$_{3}$CHO/HDCO ratio as a tool to study the COMs formation}
\includepdf[pages=-,pagecommand={\thispagestyle{plain}}]{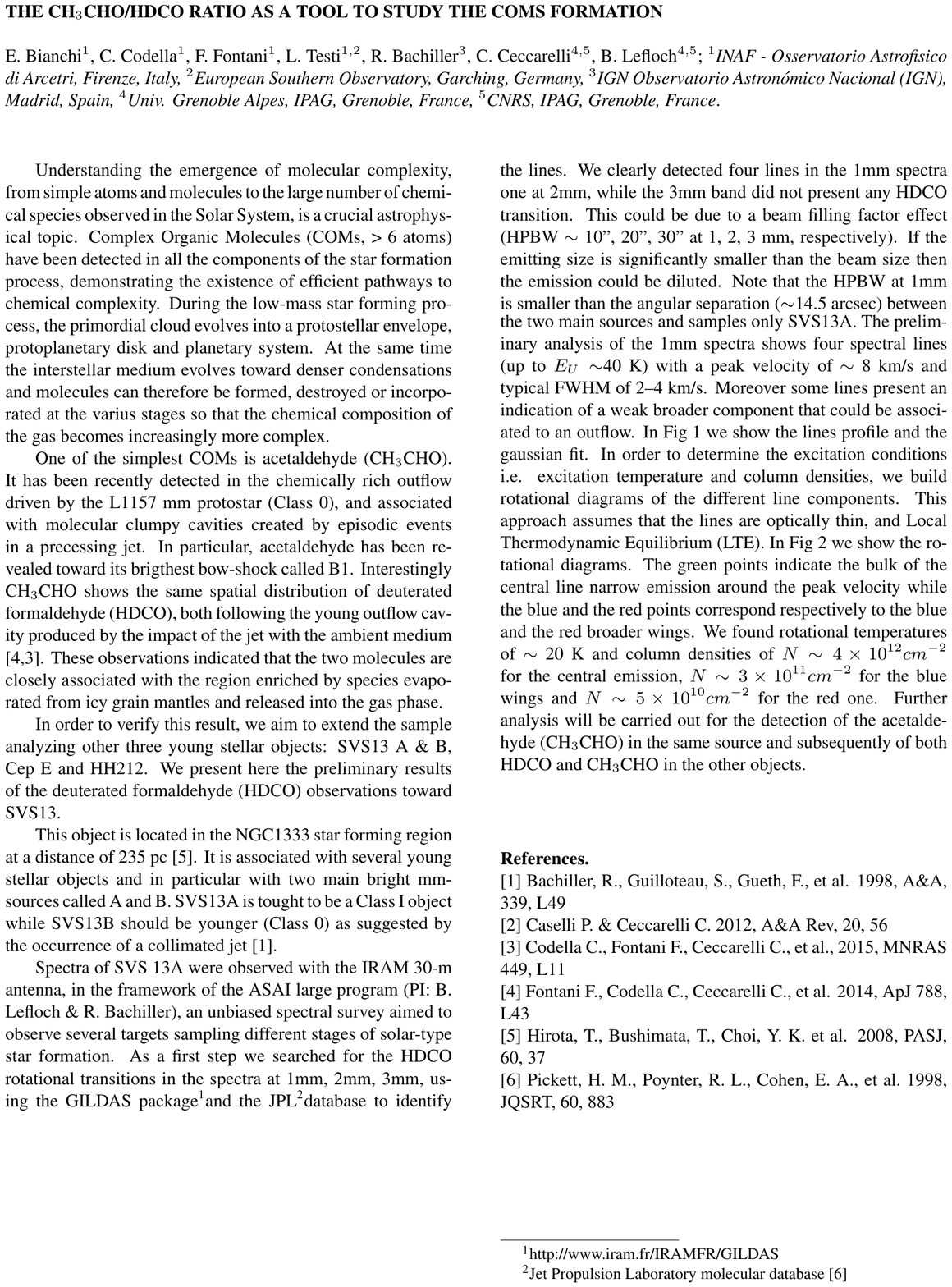}

\clearpage
\addcontentsline{toc}{subsection}{\textit{C. Codella} -- HH212's Anatomy: story of a protostellar series}
\includepdf[pages=-,pagecommand={\thispagestyle{plain}}]{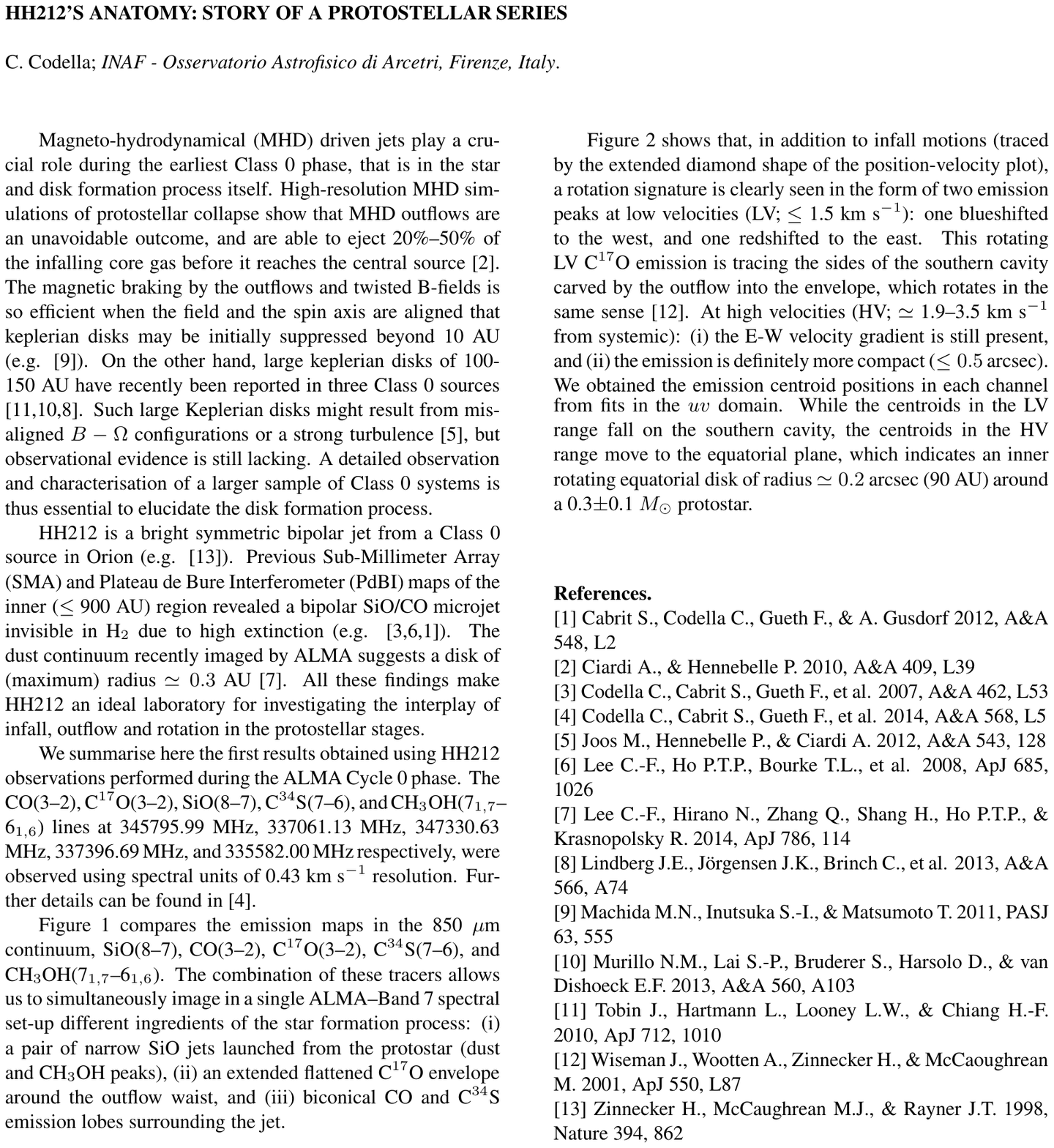}

\clearpage
\addcontentsline{toc}{subsection}{\textit{F. Fontani} -- Deuterated molecules: a chemical filter for shocks}
\includepdf[pages=-,pagecommand={\thispagestyle{plain}}]{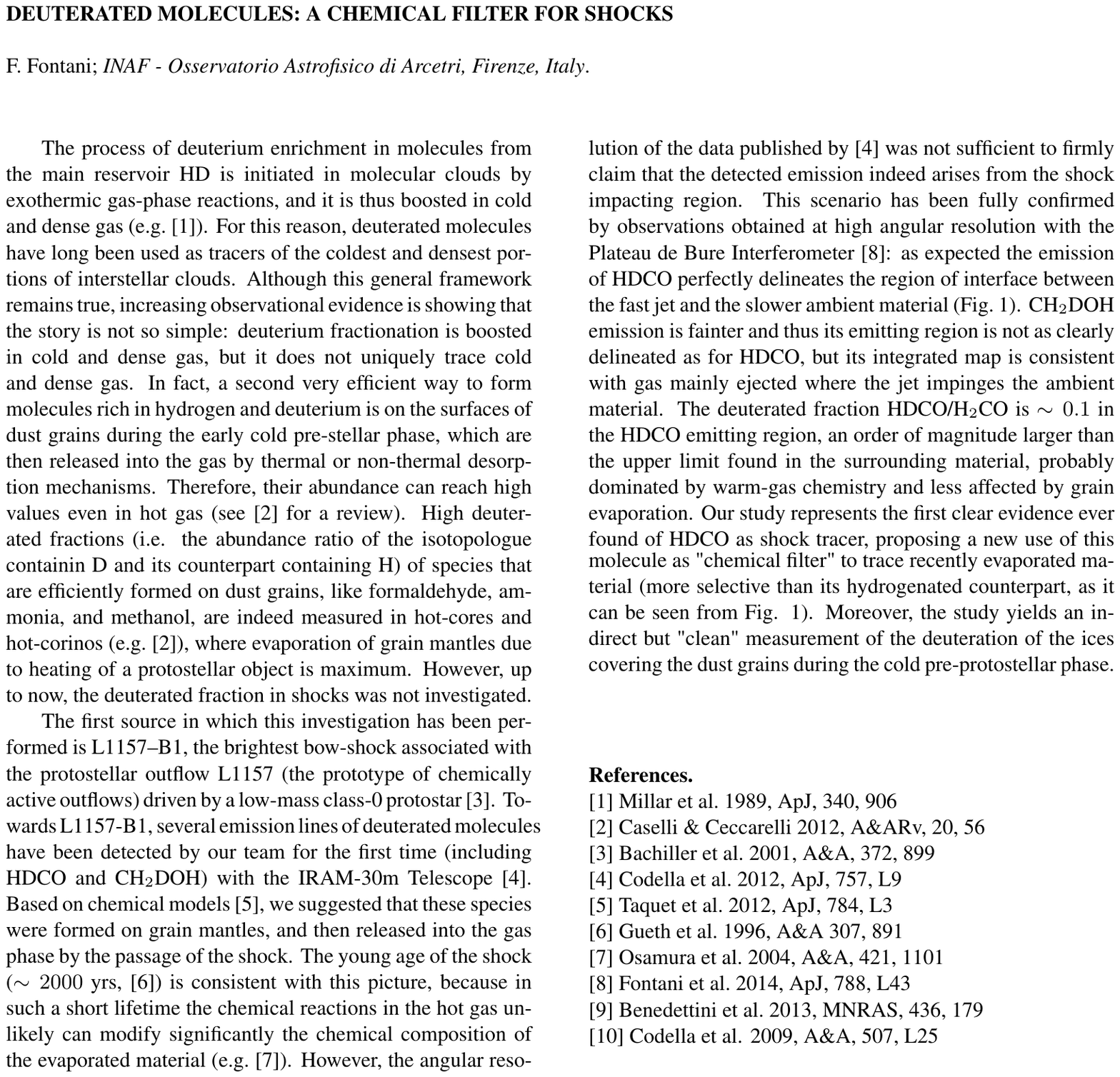}

\section*{\huge Part 2: \\~ \\ \textit{Disk Accretion and YSOs}}
\addcontentsline{toc}{section}{Part 2: Disk Accretion and YSOs}

\clearpage
\addcontentsline{toc}{subsection}{\textit{J. M. Alcal\'a} -- Search for accreting planets in transitional discs}
\includepdf[pages=-,pagecommand={\thispagestyle{plain}}]{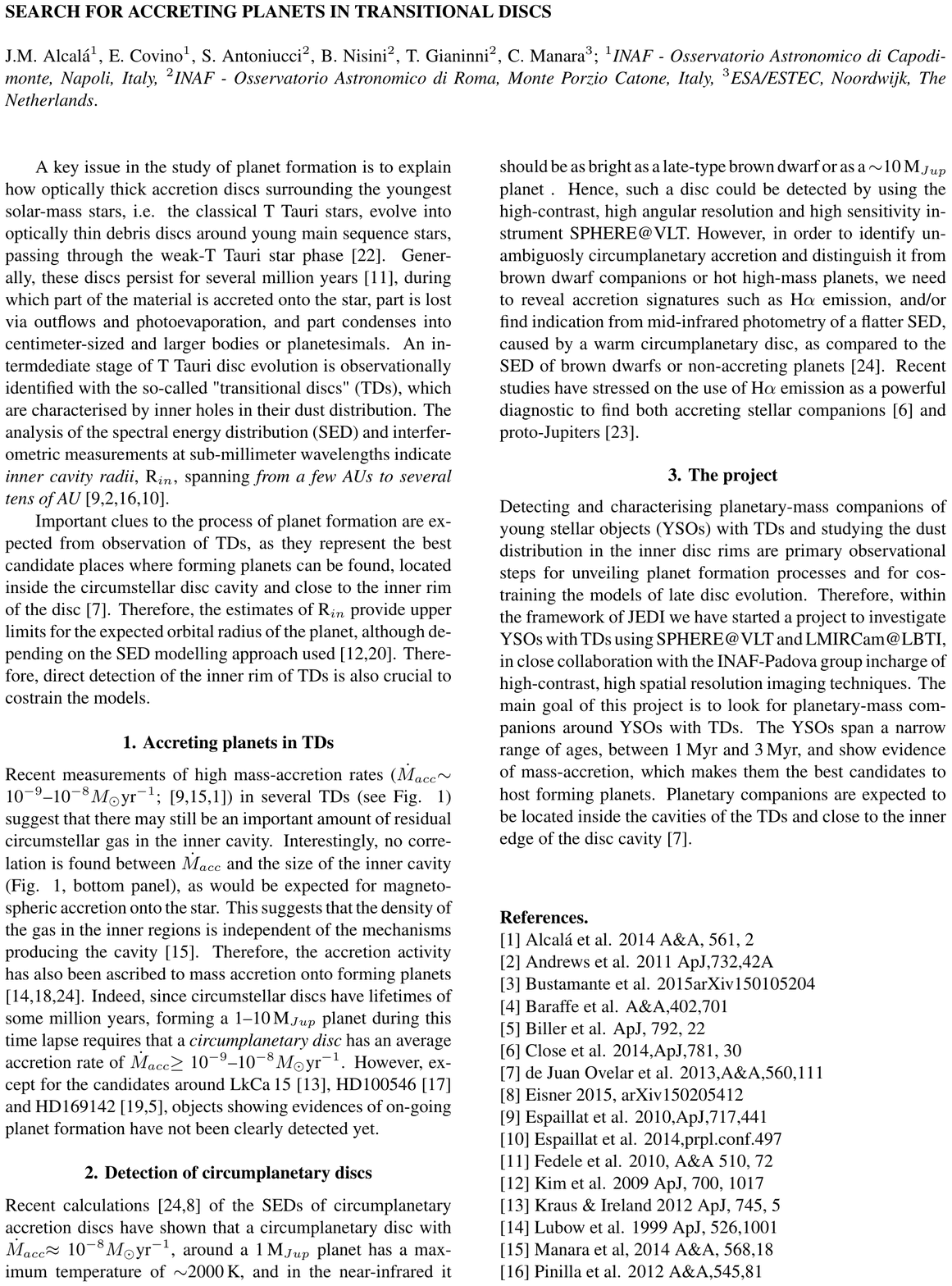}

\clearpage
\addcontentsline{toc}{subsection}{\textit{S. Antoniucci} -- HI decrements and line profiles in young stellar objects}
\includepdf[pages=-,pagecommand={\thispagestyle{plain}}]{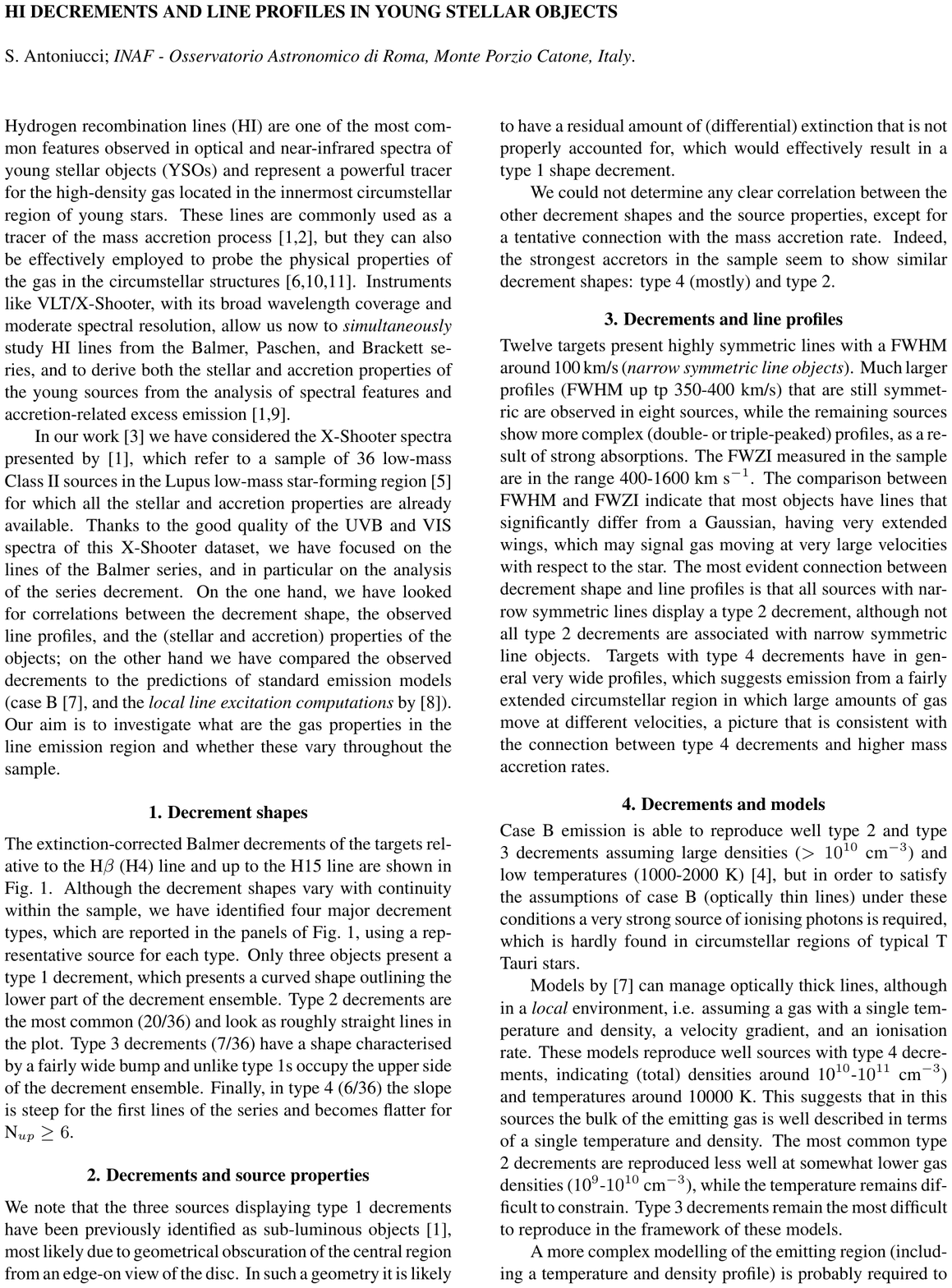}

\clearpage
\addcontentsline{toc}{subsection}{\textit{K. Biazzo} -- A study of accretion in the L1615/L1616 cometary cloud}
\includepdf[pages=-,pagecommand={\thispagestyle{plain}}]{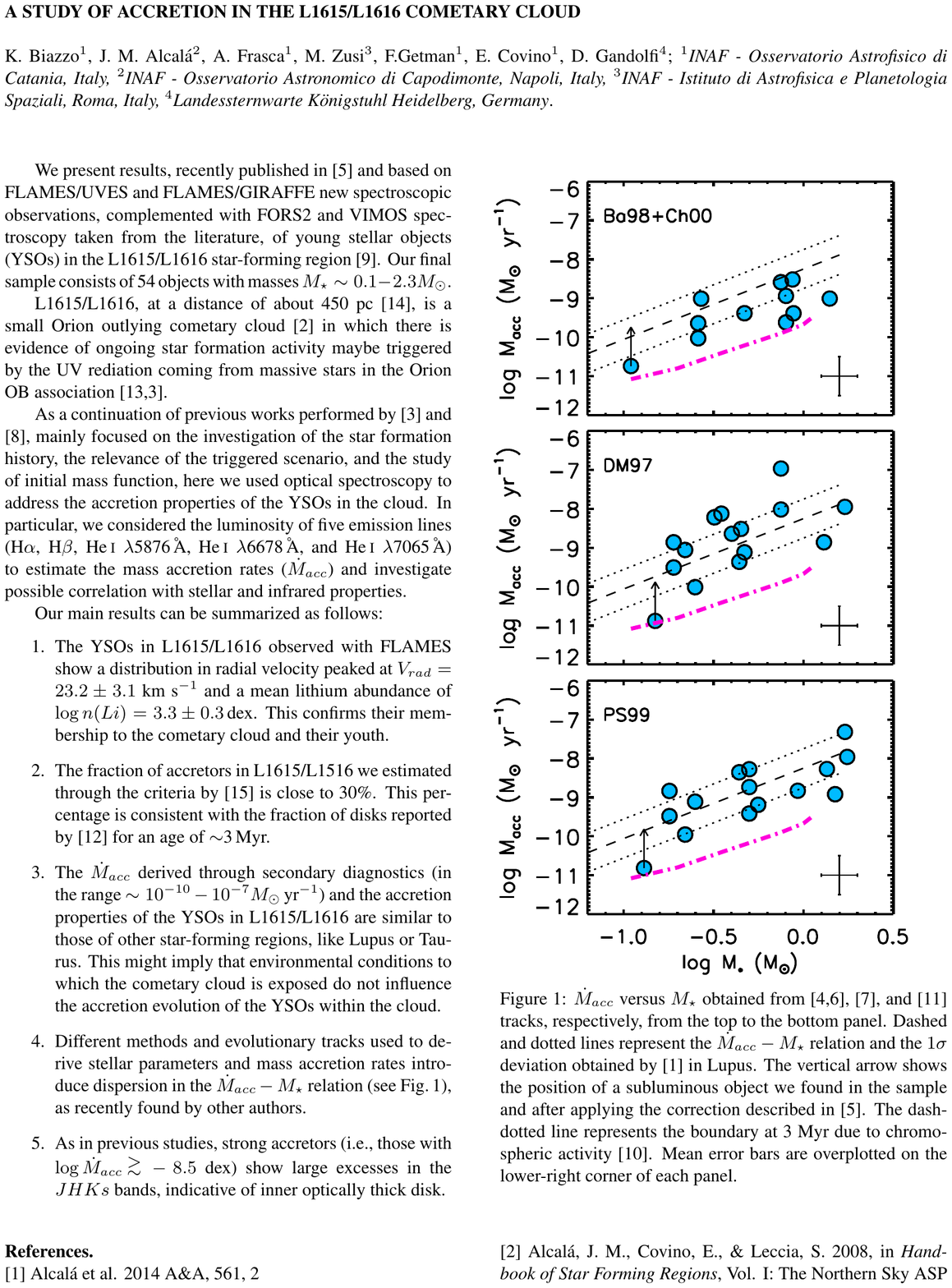}

\clearpage
\addcontentsline{toc}{subsection}{\textit{D. Fedele} -- Mass accretion in pre-main-sequence stars: the core strikes back}
\includepdf[pages=-,pagecommand={\thispagestyle{plain}}]{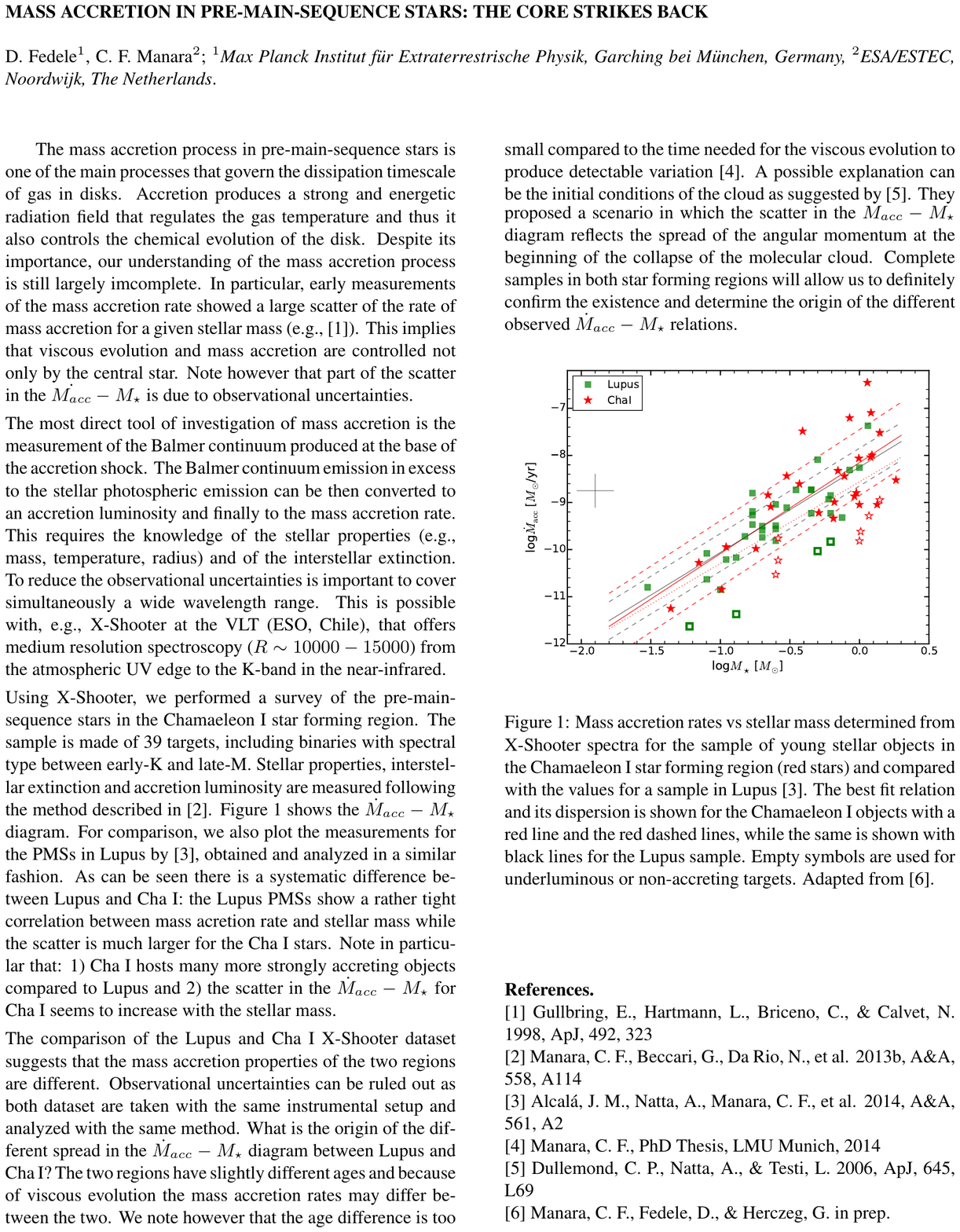}

\clearpage
\addcontentsline{toc}{subsection}{\textit{A. Frasca} -- A study of young stellar objects in the  $\gamma$~Velorum and Cha~I clusters}
\includepdf[pages=-,pagecommand={\thispagestyle{plain}}]{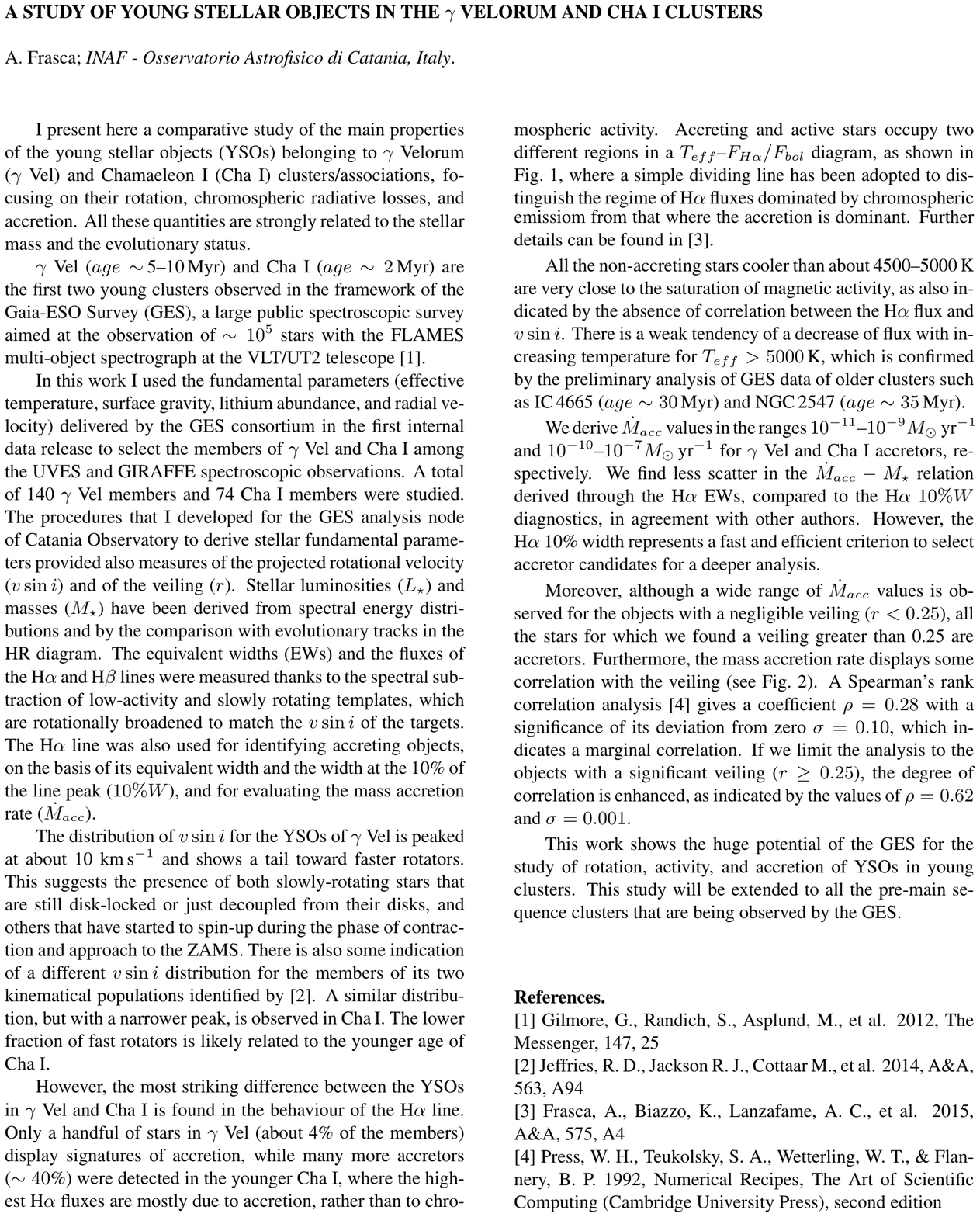}

\clearpage
\addcontentsline{toc}{subsection}{\textit{T. Giannini} -- EXORCISM: EXOR optiCal-Infrared Systematic Monitoring}
\includepdf[pages=-,pagecommand={\thispagestyle{plain}}]{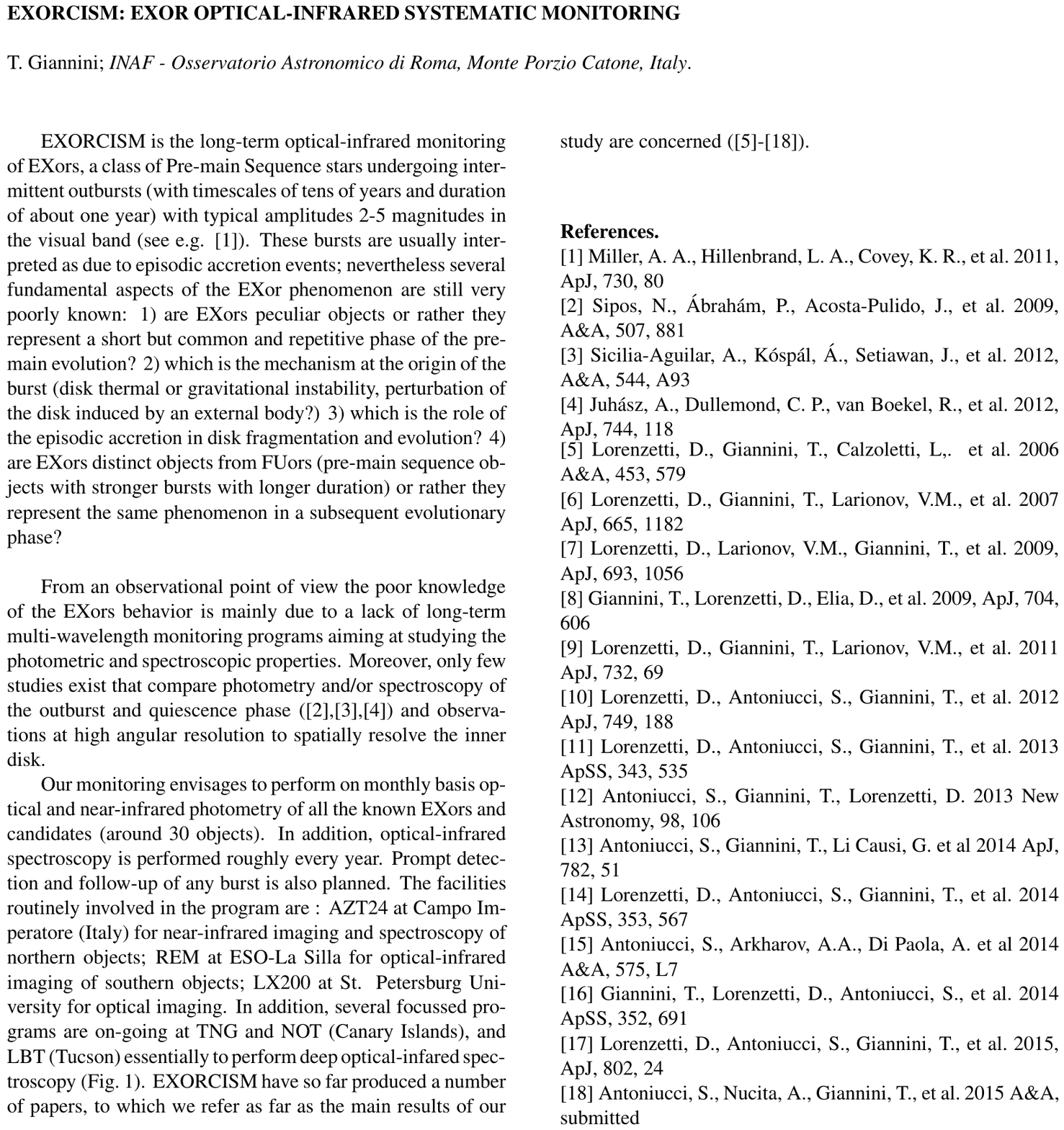}

\clearpage
\addcontentsline{toc}{subsection}{\textit{C. F. Manara} -- Accretion and wind proxies throughout disk evolution}
\includepdf[pages=-,pagecommand={\thispagestyle{plain}}]{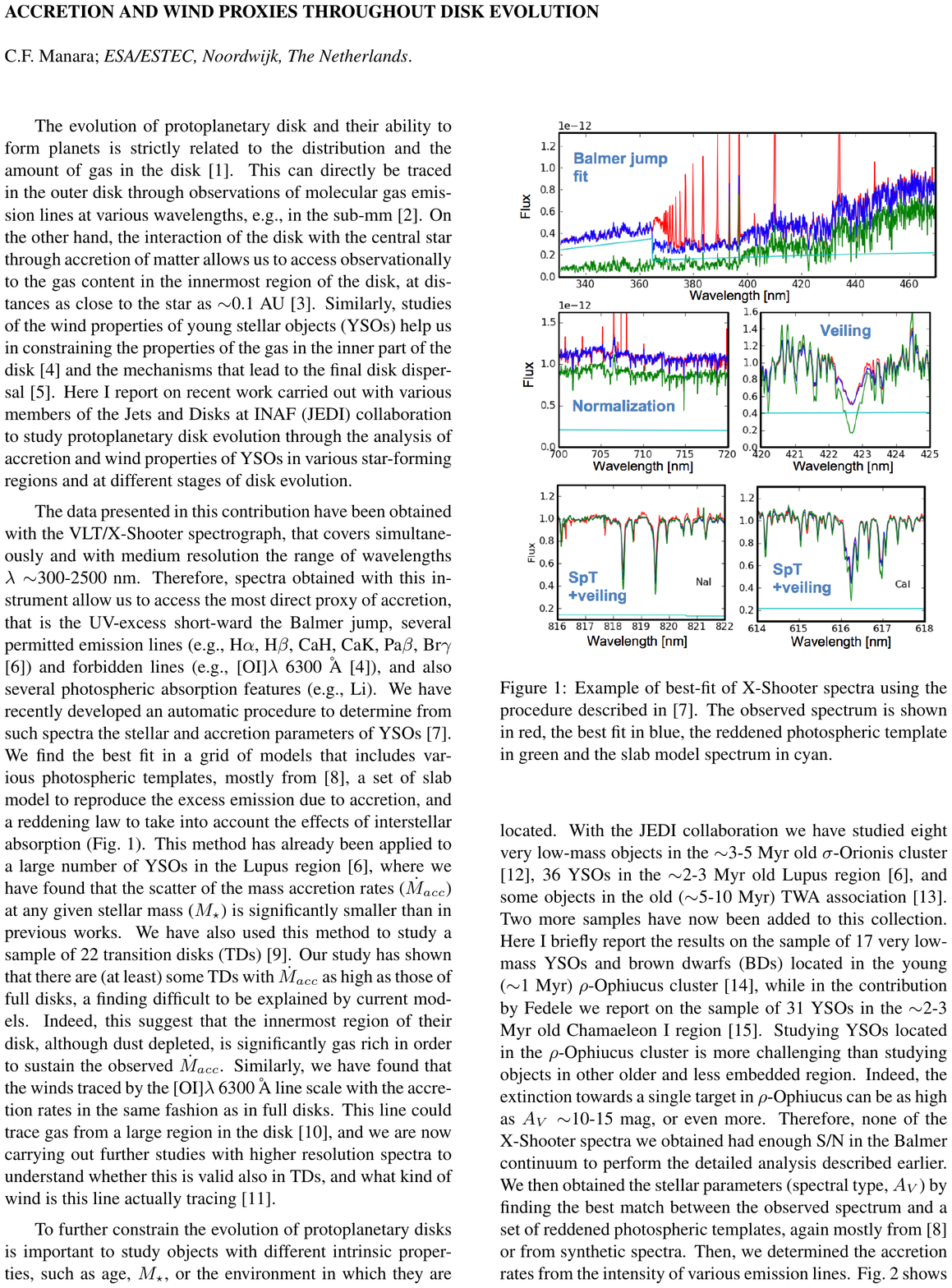}

\clearpage
\addcontentsline{toc}{subsection}{\textit{E. Rigliaco} -- Investigating Mid-Infrared Hydrogen lines as accretion indicators}
\includepdf[pages=-,pagecommand={\thispagestyle{plain}}]{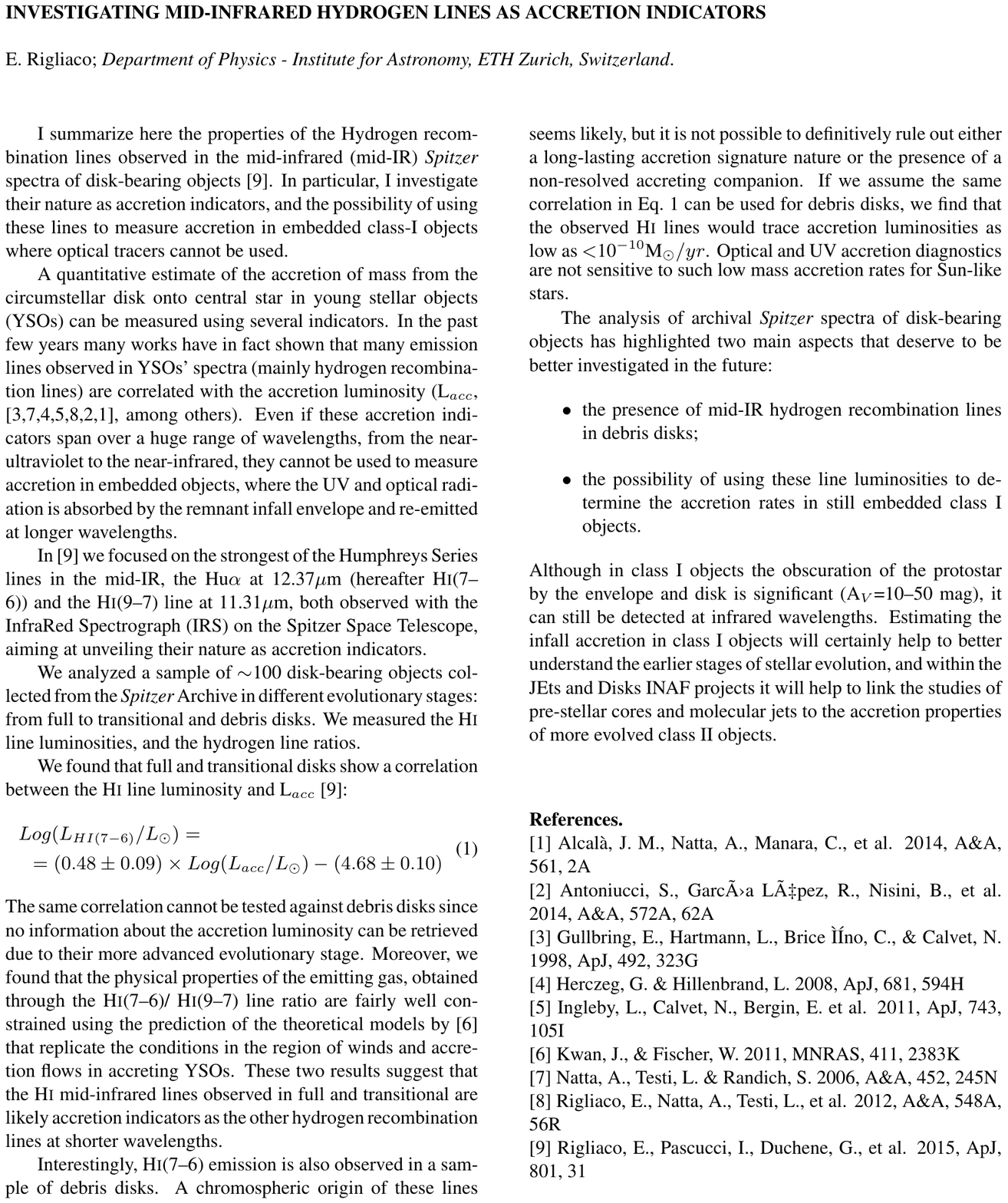}

\section*{\huge Part 3: \\~ \\ \textit{Jets and Winds}}
\addcontentsline{toc}{section}{Part 3: Jets and Winds}

\clearpage
\addcontentsline{toc}{subsection}{\textit{F. Bacciotti} -- Effects of asymmetric jets on the dynamics of protoplanetary
  disks: study of a simple model}
\includepdf[pages=-,pagecommand={\thispagestyle{plain}}]{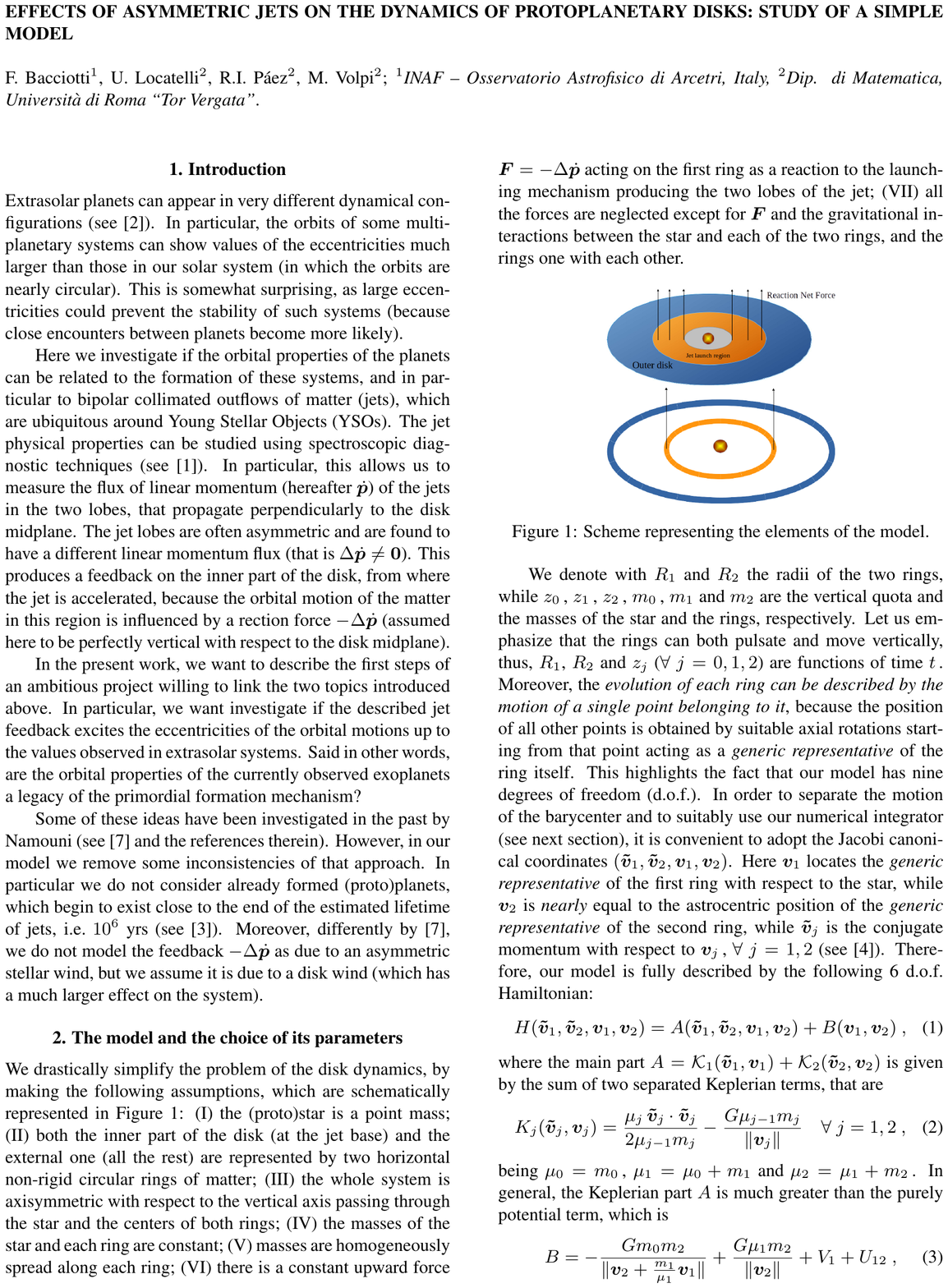}

\clearpage
\addcontentsline{toc}{subsection}{\textit{R. Bonito} -- Shocks in accretion/ejection processes: observations, models, and laboratory experiments}
\includepdf[pages=-,pagecommand={\thispagestyle{plain}}]{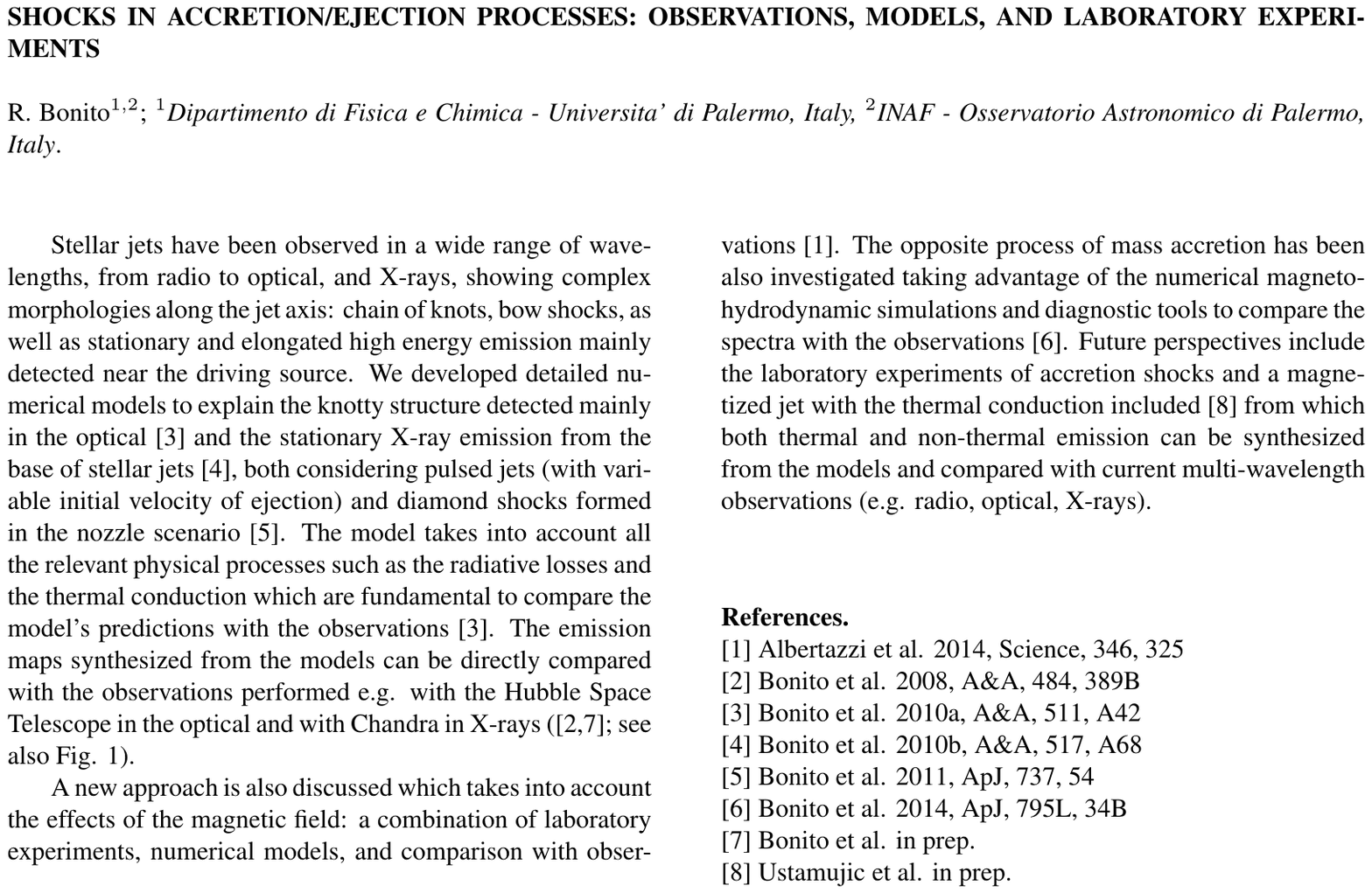}

\clearpage
\addcontentsline{toc}{subsection}{\textit{T. Giannini} -- The chemical inventory of HH1}
\includepdf[pages=-,pagecommand={\thispagestyle{plain}}]{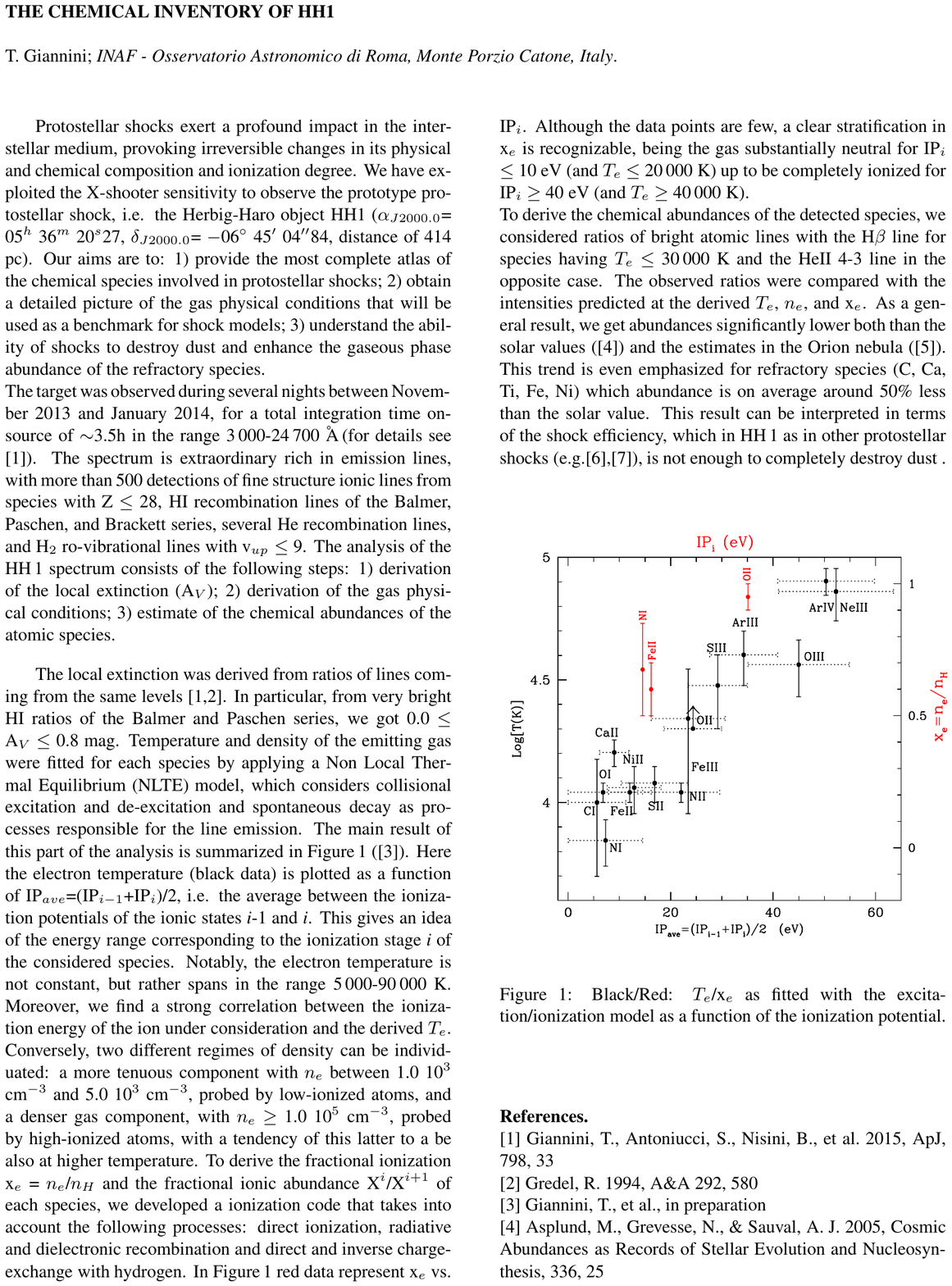}

\clearpage
\addcontentsline{toc}{subsection}{\textit{B. Nisini} -- Connection between jets, winds, and accretion in T Tauri stars}
\includepdf[pages=-,pagecommand={\thispagestyle{plain}}]{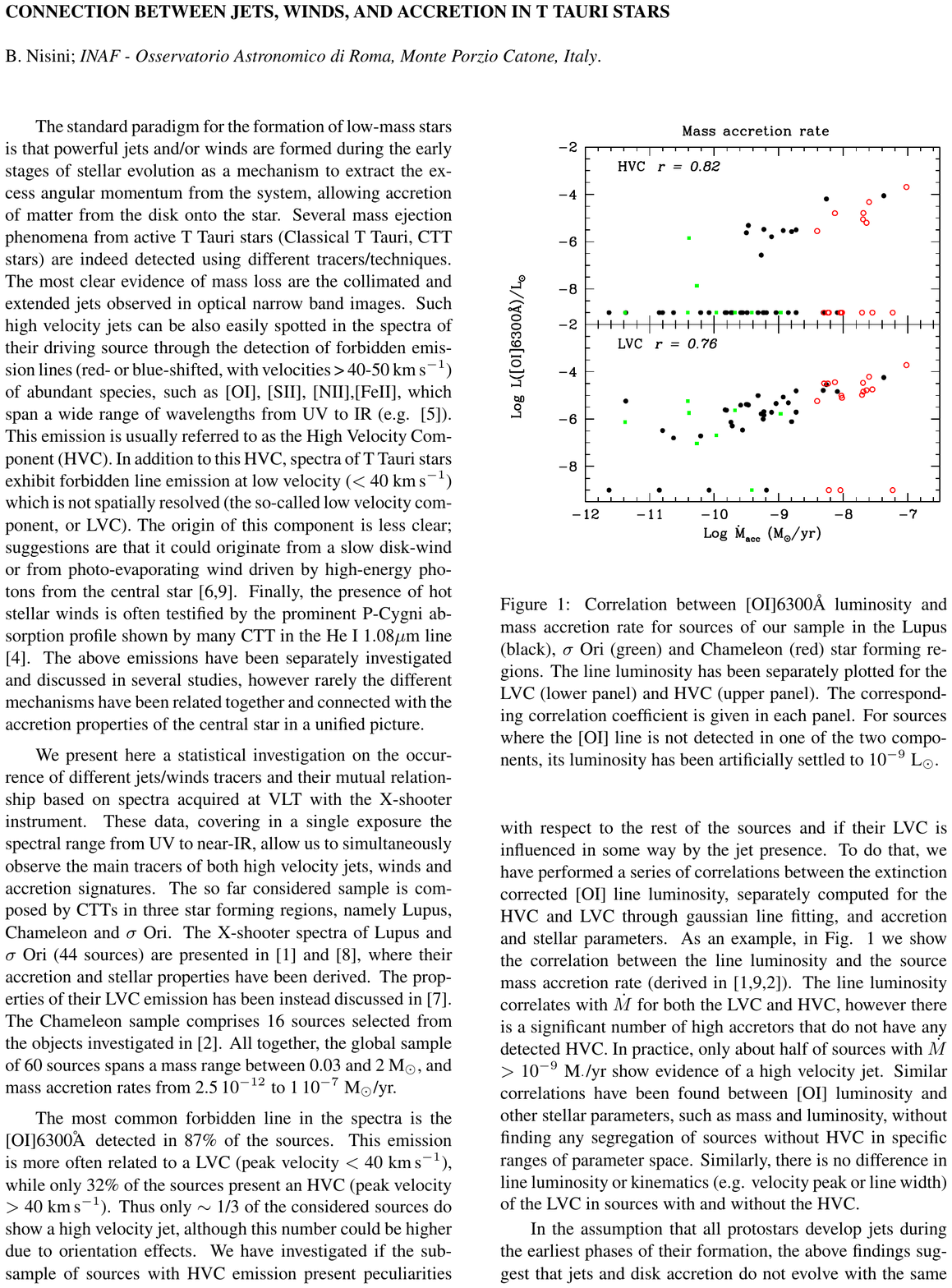}

\section*{\huge Part 4: \\~ \\ \textit{Physical and Chemical Properties of Disks}}
\addcontentsline{toc}{section}{Part 4: Physical and Chemical Properties of Disks}


\clearpage
\addcontentsline{toc}{subsection}{\textit{L. Podio} -- Disks and jets with ALMA: sculpting the birthplace of exoplanets}
\includepdf[pages=-,pagecommand={\thispagestyle{plain}}]{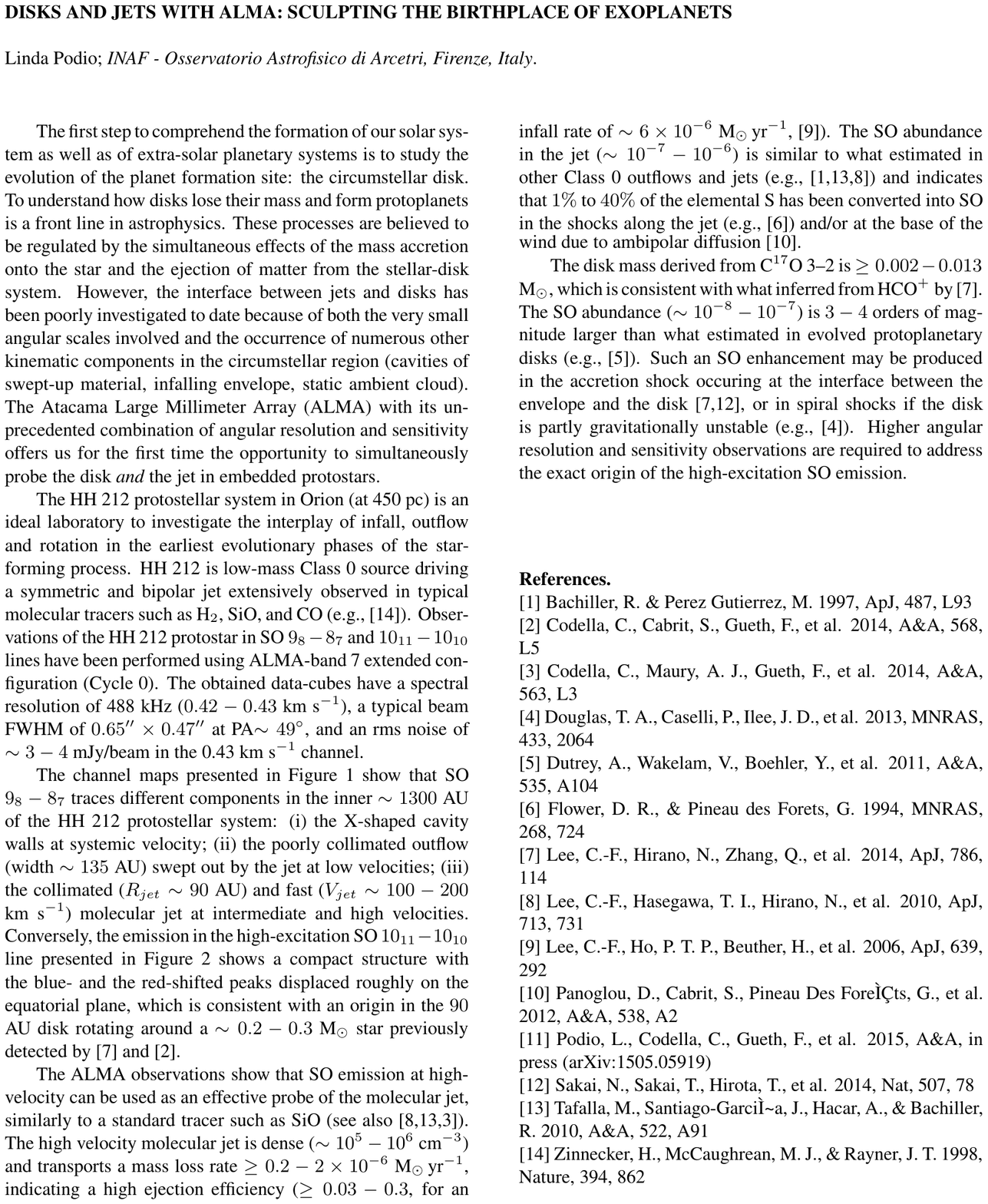}

\clearpage
\addcontentsline{toc}{subsection}{\textit{M. Tazzari} -- A new fitting tool to constrain dust grain size distribution and disk properties with ALMA and JVLA observations}
\includepdf[pages=-,pagecommand={\thispagestyle{plain}}]{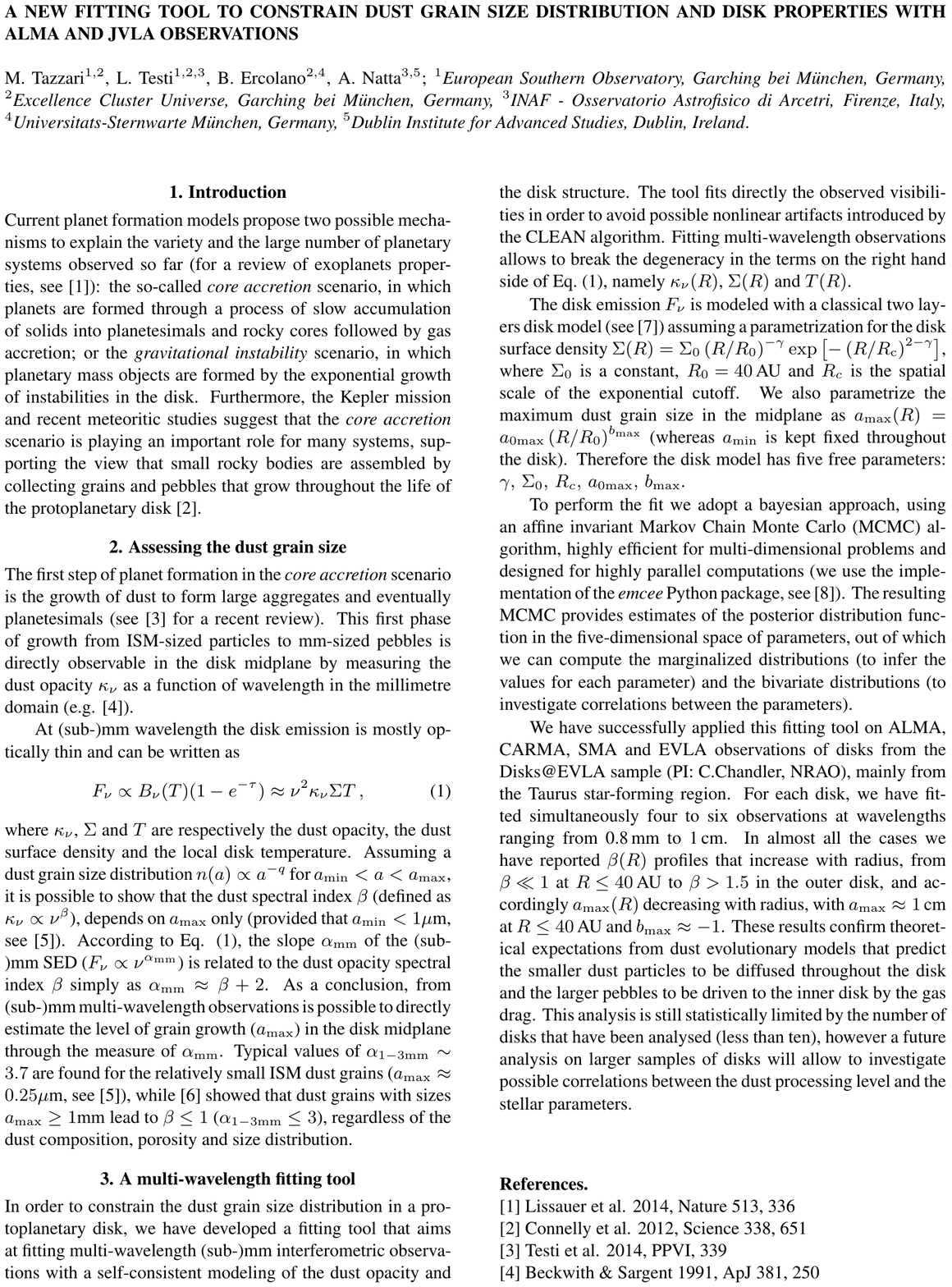}

\clearpage

%
%

\end{document}